# Colossal Seebeck coefficient of thermoelectric material calculated by space charge effect with phonon drag background

Hirofumi Kakemoto*

*Clean Energy Research Center, University of Yamanashi, 4-3-11 Takeda, Kofu, Yamanashi 400-8511, Japan*

Recently colossal Seebeck coefficient (*S*) has found in the several thermoelectric (TE) materials. We present colossal *S* and large thermal electron motivate force (EMF) reproduced by space charge (SC) model, introducing multi-Debye lengths within grain boundaries (GBs) of TE materials with phonon drag (PD) effect accompanying with electron by electron-phonon interaction. In addition to *S*, the polarity reversal was also reproduced by transfer process with inner bias around SP generated from thermal EMF. Colossal *S* and EMF for TE material were reproduced by inner SC model as a functions of averaged multi-Debye length within GBs.

## 1. Introduction

Recently, colossal Seebeck coefficient (*S*) has reported in the experimental studies about several thermoelectric (TE) materials, and it can be estimated by space charge (SC), combined with phonon drag (PD). TE properties of various TE materials have been understood in view of Fermi integral (FI) method, narrow band model [1-3] and space charge (SC) model etc. [4] Nowadays, colossal *S* with SC model defined at low temperature is firstly proposed about *n*-type $FeSb_2$ (FS) [4-8] with PD effect. [9] In addition, reversal $Cu_2Se$ (CS) [10], and *n*-type $SrTiO_3$ (STO) [1] are also paid attention for highly *S*. **Figure 1** shows TE properties of FS at around 10K, CS, and STO at around ~400K. Highly *S* was reported in FS at 10K, as shown in **Fig.1 (a)** at low temperature (~10K), and polarity reversal is reported in CS room temperature (RT) with comparing STO, as shown in **Fig.1 (b)**.

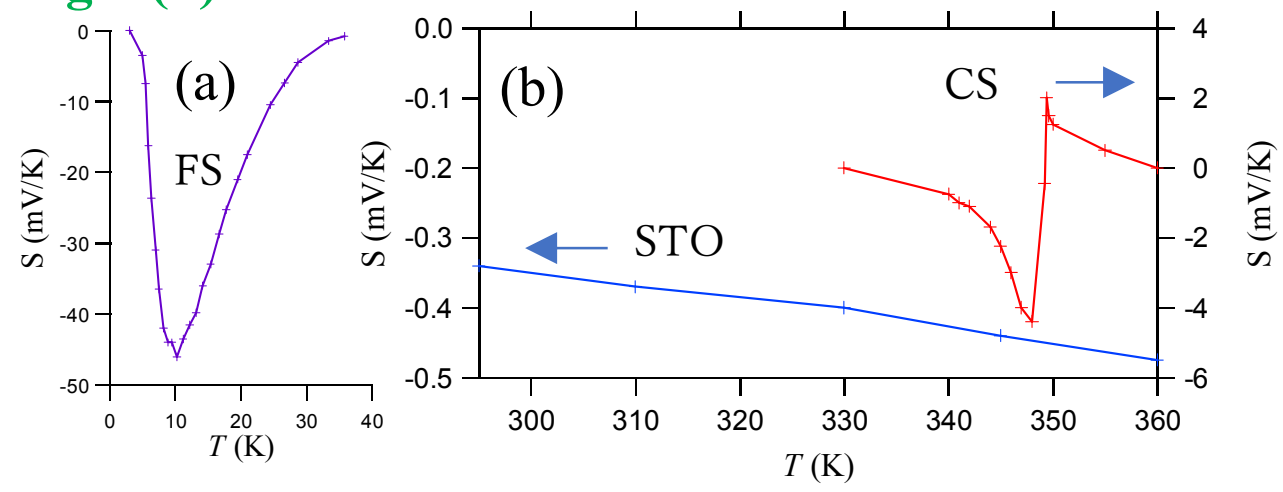

**Fig.1** TE properties of (a) *n*-type $FeSb_2$, (b) polarity reversal $Cu_2Se$ and *n*-type $SrTiO_3$ versus temperature. The polarity reversal of $Cu_2Se$ is considered caused by charge transfer process.

In this report, we report the result of reproducing colossal *S* with introducing multi-Debye length ($q_D$) inner GBs in TE material with PD effect background. [5,7]

## 2. Space charge

Space charge (SC) is not a point charge in the spatial domain, but the amount of charge generated by input is treated as a continuum of distributed charges. SC is represented by Poisson's equation as follows,

$$\nabla^2\varphi = -\rho_{SC}/\varepsilon, \quad (1)$$

eq.(1) in 1D is rewritten as, $d^2\varphi/dx^2 = -\rho_{SC}/\varepsilon_r\varepsilon_o$, (1b)

where, $\varphi$, $\rho_{SC}$, and $\varepsilon$ are potential (=$eV$), charge density of SC, and dielectric permittivity ($\varepsilon_r\varepsilon_o$), respectively. The energy of SC is also expressed as, $E=(1/2)\int\rho_{SC}\varphi dV=(1/2)\int(\mathrm{div}D)\varphi dV=k_BT_e$, where $\varphi$, and $k_BT_e$ equal $eV$, and thermal energy of electron, respectively. The estimated voltage ($V$=EMF=$SdT$) in 1D is expressed as,

$$V=(-\rho_{SC}/2\varepsilon_r\varepsilon_o)x^2+(V_o/l+\rho_{SC}l/2\varepsilon_r\varepsilon_o)x, \quad (2)$$

where $V_o$, and $l$, are inner bias, and sample length, respectively. (The electric field: $-grad(V)=-V_o/l-(\rho_{SC}l/\varepsilon_r\varepsilon_o)(1/2-x/l)$.)

The result of SC in 1D model is plotted in **Fig.2** as following eq.(2), (tentatively, within $\lambda_D$). As shown in **Fig.2**, $V$ follows as parabolic: $(-\rho_{SC}/2\varepsilon_r\varepsilon_o)x^2$ in $\varepsilon_r$ below 500 (semiconductor, or metallic, and/or vacuum), and presents to be linear: $(V_o/l+\rho_{SC}l/2\varepsilon_r\varepsilon_o)x$ up to 500 (insulator), therefore $|S|dT$ in insulator equals $V_o/l+\rho_{SC}l/2\varepsilon_r\varepsilon_o$. In **Fig.2**, $|S|dT$ is estimated as the function of $V_o$ (0-0.06 V), and $\rho_{SC}$ (1-10 μC cm$^{-1}$).

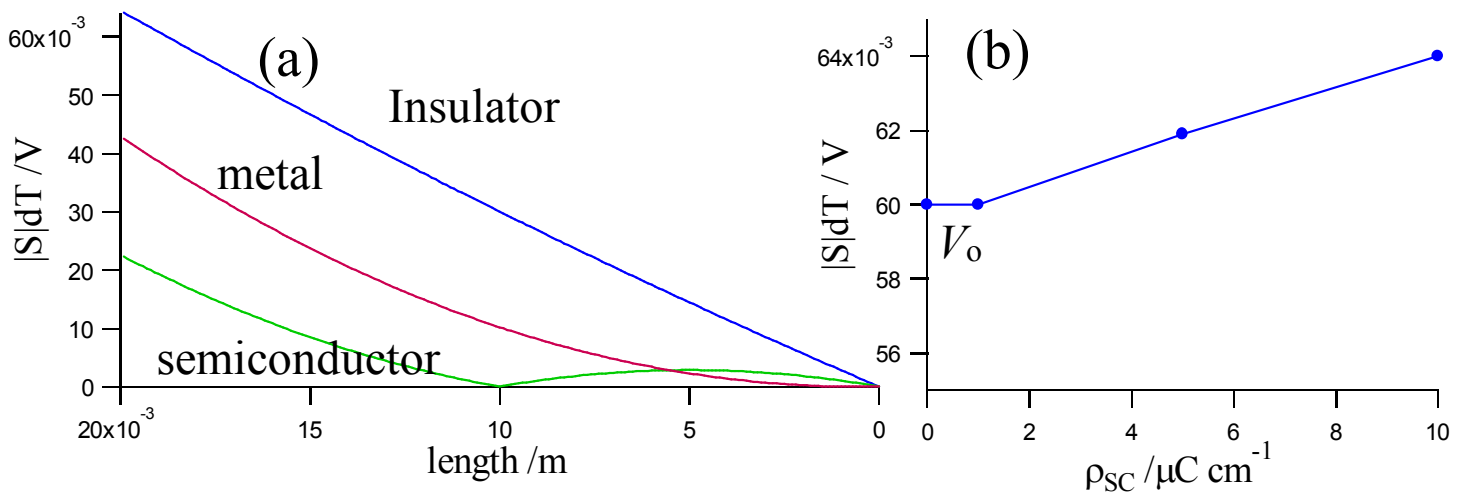

**Fig.2** (a) Estimated voltage as space charge distribution versus length, metal ($\rho_{SC}$, $\varepsilon_r$, $V_o$)=(1.0 μC cm$^{-1}$,10, 0V), semiconductor (10 μC cm$^{-1}$, 100, 0V), and insulator (10 μC cm$^{-1}$, 500, 0.06V), (b) $|S|dT$ versus $\rho_{SC}$. $|S|dT$ increases with increasing $\rho_{SC}$ from $V_o$.

In general, $\rho_{SC}$ equals $-q[N(\boldsymbol{r})-N_o]$, and $N(\boldsymbol{r})-N_o=3N_oeV(\boldsymbol{r})/2E_F$, where $E_F(\boldsymbol{r})=E_F{}^o+2E_F{}^o[N(\boldsymbol{r})-N_o]/3N_o$, therefore $\nabla^2\varphi$ equals $\lambda^{-2}V(\boldsymbol{r})$, and $V(\boldsymbol{r})=q\exp(-r/\lambda_{TF})/(4\pi\varepsilon_r\varepsilon_o)r$ where $\lambda_{TF}=[2\varepsilon_r\varepsilon_oE_F{}^o/3N_oe^2]$ (Thomas-Fermi screening, for metal). On the other hand, $N(\boldsymbol{r})-N_o=N_oq\phi(\boldsymbol{r})/k_BT$, eventually $V(\boldsymbol{r})=q\exp(-r/\lambda_D)/(4\pi\varepsilon_r\varepsilon_o)r$ where $\lambda_D=1/q_D=(\varepsilon_r\varepsilon_ok_BT_e/e^2N_e{}^2)^{0.5}$ (Debye length, for semiconductor or insulator).

## 3.Electron phonon interaction

The deformation potential for electron phonon (e-p) interaction is represented as,

$$H'=H_{e\text{-}p}=\Sigma D_qe^{iqr}\xi_q, \quad (3)$$

where, $D_q$ is the coupling constant, and $\xi_q$ is the matrix element. As shown in Fig.3, from difference of energies: $E_i-E_a=\varepsilon_{k'}-\varepsilon_{k'-q}-\hbar\omega_q$, and $E_i-E_b=\varepsilon_k-\varepsilon_{k+q}-\hbar\omega_{-q}$, $\Delta E_{ph}$ are expressed during absorption $(+\hbar\omega_q)$ or emission $(-\hbar\omega_q)$ of phonon energy for electron [6], as follows,

$$\Delta E_{ph}=|W_q|^2(1/(\varepsilon_{k'}-\varepsilon_{k'-q}-\hbar\omega_q)+1/(\varepsilon_k-\varepsilon_{k+q}-\hbar\omega_{-q}))$$
$$=|W_q|^2(1/(\varepsilon_{k+q}-\varepsilon_k-\hbar\omega_q)+1/-(\varepsilon_{k+q}-\varepsilon_k)-\hbar\omega_q))$$
$$=2|W_q|^2\hbar\omega_q/\{(\varepsilon_{k+q}-\varepsilon_k)^2-\hbar^2\omega_q{}^2\}$$
$$=-2|W_q|^2/\hbar\omega_q, \quad (4)$$

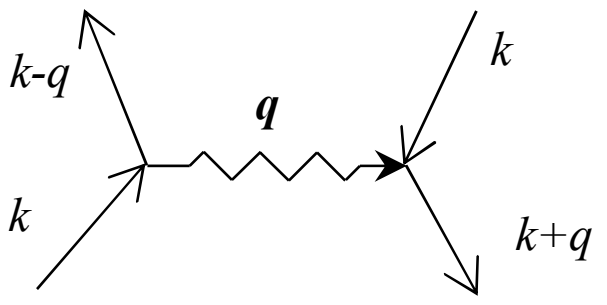

Fig.3 The illustration of electron phonon interaction.

## 4.Phonon drag effect

In eq.(3), the matrix element of $\xi_q$ is expressed as,

$$<N_q|\xi_q{}^*|N_{q-1}>=<N_{q-1}|\xi_q|N_q>=(\hbar N_q/2M\omega_qN)^{1/2}, \quad (5)$$

where, $M$ is mass of each lattice point, and $N$ is the number density of lattice point.

$$|<\psi_1|H_{e\text{-}p}|\psi_2>|^2=|W_{12}(\Delta)|^2$$
$$=(1/4)(D^2\Delta_{12}/NMs^2)J_{12}{}^2(1-\cos\boldsymbol{qR}_{12}), \quad (6)$$

where, $J_{12}$, and $R_{12}$ are overlap integral, and position (1-2).

The distribution function $(f_n)$ of carrier is represented approximately by Fermi-Dirac function: $f(\varepsilon_n)=[\exp(\varepsilon_n)+1]^{-1}$, $\varepsilon_n=(\varepsilon_n-\xi)/k_BT$, where $\xi$ is the chemical potential: $\Sigma f(\varepsilon_n)=N_{maj}-N_{min}$.

The phonon drag (PD) part of detailed balance is represented by Boltzmann eq. as follows,[5,7]

$$(\partial f/\partial t)_{pd}=(2\pi/\hbar)\int(d^3q/(2\pi)^3)|W_q|^2|f_{l,n}-f_{m,n}|\exp(-2R_{lm}/a_B)$$
$$\times\delta N_q\{\delta[E_i-E_f-\hbar\omega_q]+\delta[E_i-E_f+\hbar\omega_q]\}, \quad (7)$$

where $N_q$ equal $[\exp(\hbar\omega_q/k_BT)-1]^{-1}$, and $a_B$ is Bohr radius.

In addition to detailed balance, Seebeck coefficient caused by PD effect $(S_{pd})$ is represented as the functions of length $(l_k, l_s)$, as follows,

$$S_{pd}=C\beta\, v_a\, l_s/(\mu)T, \quad (8)$$
$$l_k=\beta\, l_s, \quad (9)$$

where $C$, $\beta$ and $\mu$ are coefficients ($0.17\text{x}10^{+3}$~$3.5\text{x}10^{+3}$, ~0.9), and drift mobility ($\mu$: $1\text{x}10^{-3}$ $m^2$/Vsec), respectively.

The statement of detailed balance for hopping (forward, or backward) is also expressed as follows, [5]

$$(\partial f/\partial t)_{lm}=(2\pi/\hbar)\int(d^3q/(2\pi)^3)|W_q|^2|\Lambda_f-\Lambda_b|\exp(-2R_{lm}/a_D), \quad (10)$$

where, forward hopping: $\Lambda_f=f_{l,n}(1-f_{m,n})\{(N_q+1)\}\delta[E_i-E_f-\hbar\omega_q]+N_q\delta[E_i-E_f+\hbar\omega_q]\}$, and backward hopping: $\Lambda_b=f_{m,n}(1-f_{l,n})\{(N_q+1)\}\delta[E_i-E_f-\hbar\omega_q]+N_q\delta[E_f-E_i+\hbar\omega_q]\}$

## 5.Space charge model(s)

The band SC models are shown in Fig.4. [5-7] In SC models, the carriers were distributed as thermally diffused for an electrode ($T_H$ or $T_C$) within grain boundaries (GBs).

In calculation, the interface SC were firstly calculated based on model I, then inner SC state was calculated based on model II.

The interface SC effect as for model I, was estimated as follows,

$$V=(3/2)^{4/3}(-j/A)^{2/3}(m^*/2e)^{1/3}x^{4/3}+SdT, \quad (11)$$

where carrier charging around electrode (1st term), and usual thermal EMF (2nd term ~0), as follows $V=V_{ele}+SdT$~$V_{ele}$, where $j=\sigma(T)E$, $j/A$, and, $m^*$ relating with temperature dependence of mobility $\mu$~$(m^*/m_o)T^{-3/2}$, are current, current density: $10^{-6}$~$10^{-3}$A/$m^2$, and $0.1m_o$~$2m_o$, respectively.

Thermal EMF by SC was also calculated by using Poisson's eq. (1b), as for SC model II, as shown in Fig.3(b). Thermal EMF and $S$ were estimated, as follows,

$$V_1=|V(l)/V(0)|$$
$$=\Sigma(\mathcal{E}/q_D)[q_Dl(1+\exp(-q_Dl))+2\exp(-q_Dl)-2], \quad (12)$$
$$S=-(V/dT)\{f(q_Dl)\}, \quad (13)$$

where $f(q_Dl)$ equals $1-(2/q_DL)\tanh(q_Dl/2)$. $N$, $\mathcal{E}$, $q_D$, and $l$ are carrier density, $eA/\varepsilon_o\varepsilon_rq_D$ at each GBs, Debye length ($\lambda_D=1/q_D$) inner GB as temperature variation, and sample length, respectively. In calculation, $q_D$ versus $T$ with GBs and EMF versus length were estimated.

The single $\lambda_D=1/q_D=(\varepsilon_r\varepsilon_ok_BT/e^2N_e)^{0.5}$~$7.43\text{x}10^{-7}\{T_e/(N_e/10^{20})\}^{0.5}$, where Te and Ne are temperature of electron (in eV) and carrier density, respectively. $\lambda_D$ and $q_Dl$ (length: $l$=100μm) are estimated as 0.2 cm, and 50 mm orders, respectively,

In addition, the multi-$q_D$ was set at GBs for series connection. [7]

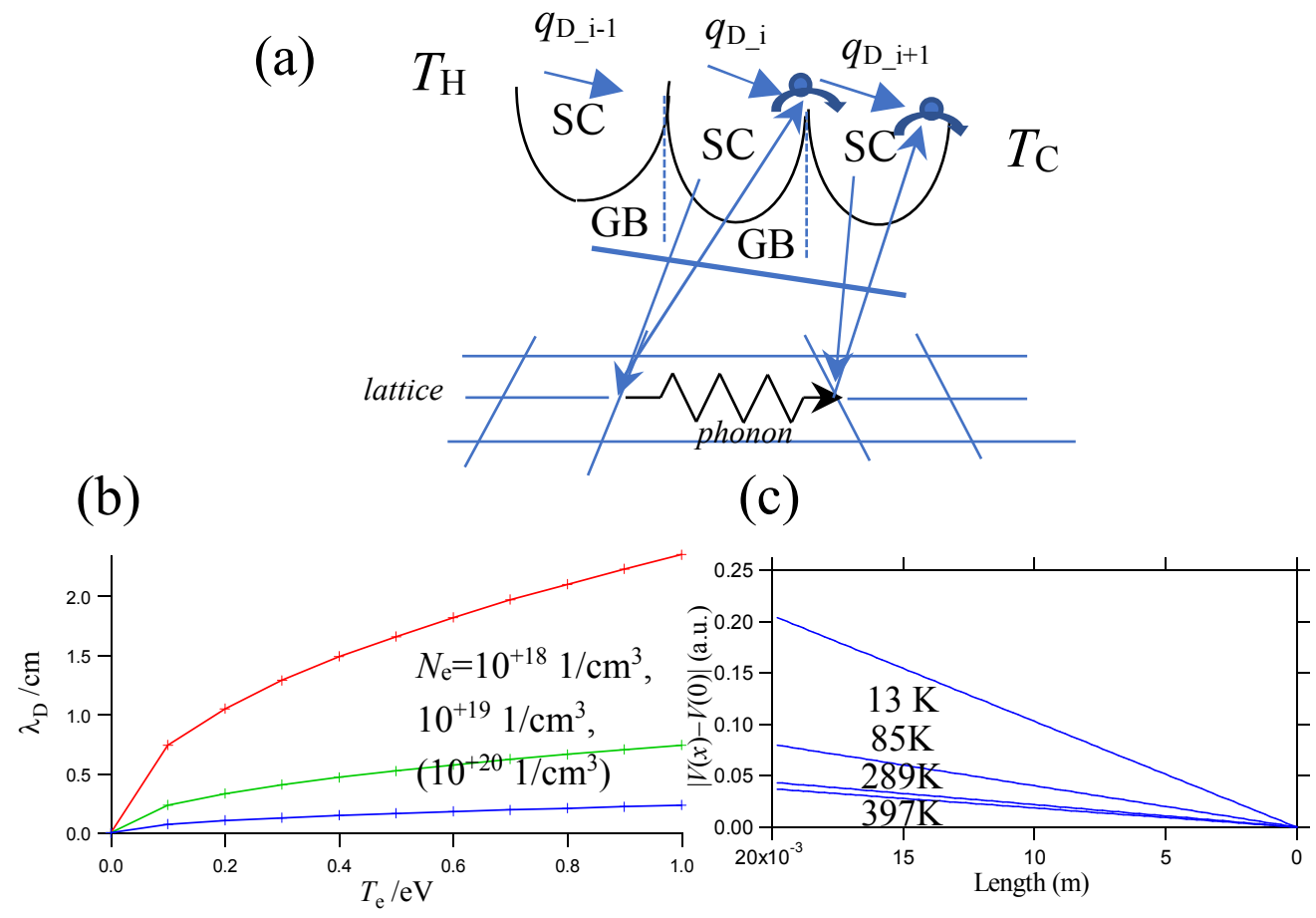

**Fig.4** (a) SC model with potential barrier of grain boundary (GB) assisted for hopping by phonon (e-p interaction), inner space charge (SC) model II, (b) Debye length versus $T_e$ as a function of $N_e$, and (c) voltage versus length as a function of $T$.

## 3.Results and discussion

### 3-1 Polarity reversal

**Figure 5** shows polarity reversal. The polarity reversal of EMF will be caused with increasing electric field of SC greater than that of prior EMF, and shorting effective length for electric field: $x_1$, $x_2$. $S$ is considered in view of the process of increasing and decreasing voltage as a function of $m^*/m_o$.

The time dependence of thermal EMF as the functions of $m^*$, and $\tau$. Inner bias is written as

$$V=E(eE/2m^*)(t/\tau)^2. \qquad (14)$$

From eq. (14): $m^*v^2/2=eE(eE/2m^*)(t/\tau)^2=eEx=eV=k_BT$, where $\tau$ is life time. Polarity reversal is caused by carrier transfer of CS, and formation of SC.

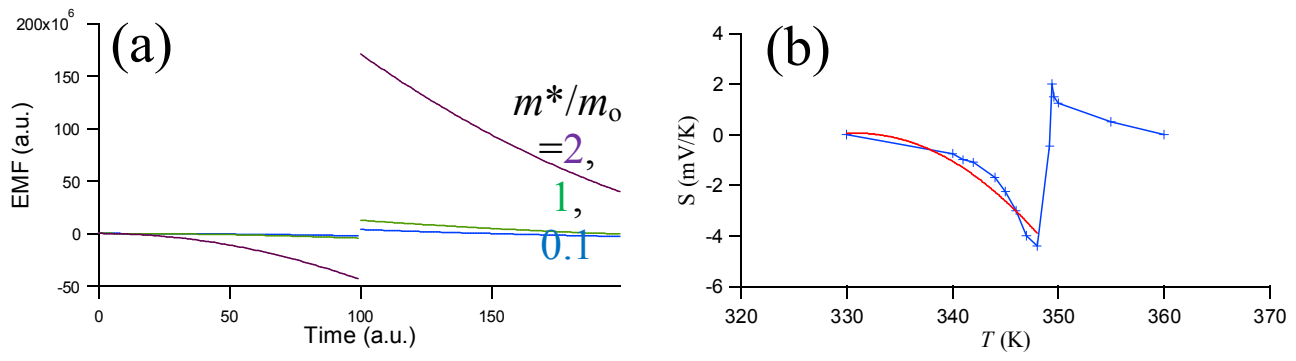

**Fig.5** Polarity of thermal EMF versus (a) time ($t$, calculation: $m^*/m_o$ purple:2, green:1, blue:0.1), and (b) $T$ of CS fitted by eq.(4) from SP model I (blue-line: exp.[8], red-line: calculation).

### 3.2 SC model I

**Figure 6(a)** shows length dependence of thermal EMF and $n$-type $S$ in TE material as a function of $m^*/m$ calculated by using eq.(11). Following SC model I in **Fig. 6(a)**, input to 1.0x10$^{-3}$ A/m$^2$ (1.0x10$^{-7}$ A/cm$^2$), 1.0x10$^{-6}$ A/m$^2$ (1.0x10$^{-10}$ A/cm$^2$), $\varepsilon_r$=10, $m^*/m_o$=0.1, 1, 2, and $dT$=10K. The profile of thermal EMF in **Fig.6(a)** is similar with that of SP model I calculation in **Fig.4(c)**. The inner electric field in SC region(s) is about 1.6x10$^{+3}$ V/m. From 0 m to 0.5 x 10$^{-2}$ m, SC is charged with inner electric field.

### 3.3 SC model II with GBs

**Figure 6(b)** shows length dependence of EMF and $S$ by using eq.(13). The voltage and $p$-type $S$ profiles are calculated by eq.(13) using $q_D$. $\lambda_D=1/q_D$ is set to 0.2 cm at GBs (grain size: 100 $\mu$m), by $q_D$ with $N_e$=10$^{+18}$~10$^{+19}$ cm$^{-3}$, and $\varepsilon_r$=10.

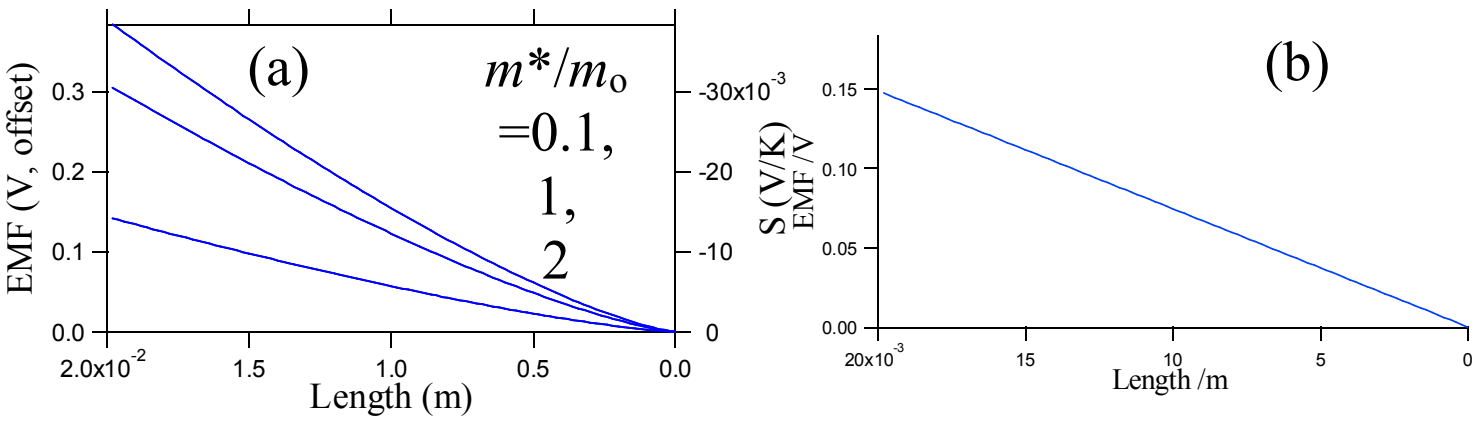

**Fig.6** Thermal EMF and $S$ versus sample length profiles (a) calculated by using eq.(11) from model I: 1x10$^{-3}$ A/m$^2$, and (b) calculated by using eq.(13) from model II by using averaged-$q_D$.

## 4.Conclusion

Thermoelectric properties (EMF, Seebeck coefficient: $S$) were calculated by interface and inner space charge (SC) models using averaged multi-Debye length within grain boundaries (GBs), (1) SC-interface around electrode, and (2) SC-inner TE material. The inner SCs are captured within GBs, and carriers are partly thermally diffused GB to GB interacted with phonon (e-p interaction). The phonon drag, and hopping are described by Boltzmann equation. EMF and $S$ were reproduced by above SC models.

* Current address: TechnoPro R&D Fukuoka Branch, 1-10-4 Hakata Eki Minami, Hakata-ku, Fukuoka 812-0016, Japan (retired)